\documentclass[aps,prl,reprint,superscriptaddress]{revtex4-2}

\usepackage{ulem}
\usepackage{xcolor}
\usepackage[colorlinks=true]{hyperref}
\usepackage{soul}
\usepackage{booktabs}
\usepackage{mathrsfs}
\usepackage{amssymb}
\usepackage{array}

\usepackage{amsmath,amssymb,color,epsfig,graphicx}
\allowdisplaybreaks[4]

\newcommand{\be}{\begin{equation}}
\newcommand{\ee}{\end{equation}}
\newcommand{\bea}{\setlength\arraycolsep{2pt} \begin{eqnarray}}
\newcommand{\eea}{\end{eqnarray}}
\newcommand{\nn}{\nonumber}

\def\0{{\sst{(0)}}}
\def\1{{\sst{(1)}}}
\def\2{{\sst{(2)}}}
\def\3{{\sst{(3)}}}
\def\4{{\sst{(4)}}}
\def\5{{\sst{(5)}}}
\def\6{{\sst{(6)}}}
\def\7{{\sst{(7)}}}
\def\8{{\sst{(8)}}}
\def\sst#1{{\scriptscriptstyle #1}}

\newcommand{\tr}{\operatorname{tr}}

\begin{document}

\hypersetup{
    linkcolor=blue,
    citecolor=red,
    urlcolor=magenta
}


\title{The fate of hidden symmetries under magnetic Harrison transformations}




\author{Shoulong Li}
\email[]{shoulongli@hunnu.edu.cn}
\affiliation{Department of Physics, Key Laboratory of Low Dimensional Quantum Structures and Quantum Control of Ministry of Education, and Institute of Interdisciplinary Studies, Hunan Normal University, Changsha, 410081, China}
\affiliation{Hunan Research Center of the Basic Discipline for Quantum Effects and Quantum Technologies, Hunan Normal University, Changsha, 410081, China}

\author{Fangkang Yu}
\email[]{fangkangyu@hunnu.edu.cn}
\affiliation{Department of Physics, Key Laboratory of Low Dimensional Quantum Structures and Quantum Control of Ministry of Education, and Institute of Interdisciplinary Studies, Hunan Normal University, Changsha, 410081, China}
\affiliation{Hunan Research Center of the Basic Discipline for Quantum Effects and Quantum Technologies, Hunan Normal University, Changsha, 410081, China}

\author{Hong-Da Lyu}
\email[Corresponding author: ]{hongdalyu@sdu.edu.cn}
\affiliation{Key Laboratory of Particle Physics and Particle Irradiation (MOE),
Institute of Frontier and Interdisciplinary Science,
Shandong University, Qingdao, Shandong, 266237, China}


\date{\today}

\begin{abstract}
Solution-generating transformations construct new geometries from known seeds, raising the question of whether and to what extent the resulting solutions inherit the hidden symmetries underlying the seeds' integrability and separability. We have established universal perturbative inheritance laws for CKTs and massless scalar separability directly at the level of magnetic Harrison transformations, encompassing both magnetization of Carter-type seeds and demagnetization of conformal Carter-type seeds. In both magnetized Kerr and demagnetized Kerr-Bertotti-Robinson, this perturbative inheritance coexists with the loss of exact Carter-type separability, while seed-connected conformal Killing-Yano structures are already obstructed at the same perturbative order. These results reveal a hierarchy in the perturbative survival of hidden symmetries and provide a framework for investigating their inheritance in more general theoretical settings and under other solution-generating transformations.
\end{abstract}



\maketitle


\textit{Introduction.}
The Kerr metric underpins the study of rotating black holes, with much of its remarkable analytical tractability stemming from its hidden symmetries~\cite{Carter:1968rr,Walker:1970un,Carter:1979fe,Frolov:2006dqt,Frolov:2017kze}. Its Killing tensor (KT) gives rise to the Carter constant and renders geodesic motion completely integrable, while the associated separability structure also permits the separation of the scalar wave equation. The corresponding Killing-Yano (KY) two-form provides a first-order symmetry operator underlying the separability of the Dirac equation. Kerr also serves as a seed for solution-generating transformations, which exploit symmetries of the reduced field equations to construct broader families of rotating geometries~\cite{Stephani:2003tm,Geroch:1970nt,Kinnersley:1973tae}. In particular, an Ehlers transformation introduces a NUT (Newman-Unti-Tamburino) charge~\cite{Demianski:Newman:1966}, a charging Harrison transformation generates an electric charge~\cite{Harrison:1968wue}, and a magnetic Harrison transformation immerses Kerr in an external Melvin-type magnetic field~\cite{Ernst:1976mzr,Ernst:1976bsr,Gibbons:2013yq}, all within standard four-dimensional Einstein-Maxwell theory. This raises a natural question: to what extent do solution-generating transformations preserve the hidden symmetries of their seeds?

Among these examples, Kerr-NUT and Kerr-Newman retain the Carter structure of Kerr and its associated separability, whereas Kerr-Melvin is of Petrov type I and lies outside the Carter family. However, an external magnetic field does not necessarily destroy algebraic speciality. Recently, Podolsk\'y and Ovcharenko~\cite{Podolsky:2025tle,Ovcharenko:2026tos} constructed two rotating black holes immersed in a Bertotti-Robinson-type magnetic field, namely the original and genuine Kerr-Bertotti-Robinson (KBR) solutions, both of which remain of Petrov type D and are conformal to Carter spacetimes~\cite{Gray:2025lwy}. Remarkably, inverse magnetic Harrison transformations map these geometries to two Ricci-flat rotating spacetimes of Petrov type I, removing the Maxwell field while leaving its original magnetic-field parameter as a geometric deformation~\cite{Ma:2026otg,Ma:2026gdn}. These contrasting geometries provide a natural setting for investigating how hidden symmetries evolve under magnetic Harrison transformations and their inverses. In particular, previous weak-field studies have found massless scalar separability in slowly rotating magnetized Kerr~\cite{Konoplya:2008hj}, while perturbative null Hamilton-Jacobi separability also emerges in demagnetized KBR~\cite{Lyu:2026pqh}. These observations motivate a more specific question: do these perturbative separability properties follow from a general inheritance mechanism intrinsic to magnetic Harrison transformations, and does this mechanism also protect conformal Killing-Yano (CKY) structures?

In this Letter, we address these questions directly at the level of magnetic Harrison transformations, deriving general inheritance laws for conformal Killing tensors (CKTs) and massless scalar separability at the leading nontrivial order under both magnetization and demagnetization between vacuum and electrovacuum geometries. Using magnetized Kerr~\cite{Ernst:1976bsr} and demagnetized KBR~\cite{Ma:2026otg} as representative examples, we further establish coordinate-independent obstructions to exact Carter-type separability. In both cases, CKY structures are already obstructed at the same perturbative order, revealing a distinct hierarchy of perturbative inheritance in which CKTs and massless scalar separability survive while CKY symmetry does not.

\textit{Magnetic Harrison transformations.}
For a stationary solution of four-dimensional Einstein-Maxwell theory admitting a spacelike Killing vector $k=\partial_z$, the metric and gauge potential can be decomposed as
\begin{equation}
ds^2=e^{-2\varphi}(dz+\omega)^2+e^{2\varphi}ds_3^2 \,,
\quad
A=\chi(dz+\omega)+{\cal A} \,,
\label{eq:kk}
\end{equation}
where $ds_3^2$ is the metric on the three-dimensional base, $\omega$ and ${\cal A}$ are one-forms on this base. All quantities are independent of $z$. Upon dimensional reduction, the one-forms can be dualized into scalar potentials $\sigma$ and $\psi$ via
\begin{equation}
e^{-2\varphi}\ast(d{\cal A}+\chi\,d\omega)=d\psi \,,
\quad
\frac12e^{-4\varphi}\ast d\omega=d\sigma-2\chi\,d\psi \,,
\label{eq:dual}
\end{equation}
where $\ast$ denotes the Hodge dual associated with $ds_3^2$. The four scalar potentials $(\varphi,\chi,\psi,\sigma)$ parametrize an $SU(2,1)/U(2)$ sigma model~\cite{Gibbons:2013yq}. In terms of its Hermitian coset matrix ${\cal M}$, the Harrison transformation is
\begin{equation}
\bar{\cal M}=U^\dagger{\cal M}U \,,
\quad
U({\cal B})=
\begin{pmatrix}
1&0&0\\
{\cal B}/\sqrt2&1&0\\
{\cal B}^2/4&{\cal B}/\sqrt2&1
\end{pmatrix} \,,
\label{eq:harrison}
\end{equation}
where ${\cal B}$ is the transformation parameter. The forward transformation generated by $U({\cal B})$ and its inverse generated by $U(-{\cal B})=U({\cal B})^{-1}$ are referred to as magnetization and demagnetization, respectively. Since we focus on transformations relating Ricci-flat vacuum solutions to electrovacuum configurations immersed in an external magnetic field, barred and unbarred quantities denote the magnetized and vacuum configurations, respectively. The transformation~\eqref{eq:harrison} leaves $ds_3^2$ unchanged while modifying the scalar potentials. Defining $\alpha=e^{-2\varphi}+\chi^2+\psi^2$ and $\beta=\sigma-\chi\psi$, the transformed scalars are given by
\begin{align}
\bar\varphi&=\varphi+\frac12\log\gamma \,,\qquad
\bar\chi=\frac12\partial_{\mathcal B}\log\gamma \,,\nonumber\\
\bar\psi&=\left[\psi+\mathcal B\beta+\mathcal B^2(2\chi\beta-\psi\alpha)/4\right]/{\gamma} \,,\nonumber\\
\bar\sigma&=\bar\chi\bar\psi+
\left[(1+\mathcal B\chi)\beta-\mathcal B\psi\alpha/2\right]/{\gamma} \,,
\label{eq:scalarmap}
\end{align}
where $\gamma=(1+{\cal B}\chi+{\cal B}^2\alpha/4)^2+({\cal B}\psi+{\cal B}^2\beta/2)^2$. The transformed $\bar\omega$ and $\bar{\cal A}$ are then reconstructed, yielding the metric $d\bar{s}^2$.


\textit{Perturbative inheritance of CKTs and massless scalar separability.}
Starting from a fixed vacuum seed with $\chi={\cal A}=0$, Eqs.~\eqref{eq:dual} and \eqref{eq:scalarmap} yield the following expansions of the magnetized metric functions for small ${\cal B}$:
\be
\bar\varphi
=\varphi
+\frac{{\cal B}^2}{4} e^{-2\varphi}
+O({\cal B}^4) \,,
\quad
\bar\omega
=\omega +O({\cal B}^4) \,.
\label{eq:general-magnetization-expansion}
\ee
Thus, the inverse metric and determinant transform as
\bea
\bar{g}^{\mu\nu}&=&g^{\mu\nu}+{\cal B}^2\left(k^\mu k^\nu-e^{-2\varphi}g^{\mu\nu}/2\right)+{\cal O}({\cal B}^4) \,,\label{eq:inverseg}
\\
\sqrt{-\bar g}
&=&\left(1+{\cal B}^2e^{-2\varphi}/2\right)
\sqrt{-g}+{\cal O}({\cal B}^4) \,.
\label{eq:protection}
\eea
These relations are the common origin of the Hamiltonian and wave-operator protection identities below.

The Hamiltonian $H=g^{\mu\nu}p_\mu p_\nu/2$ of the vacuum geometry and its magnetized counterpart $\bar H$ satisfy
\be
\bar H=H+\frac{{\cal B}^2}{2}\left(J^2-e^{-2\varphi}H\right)+{\cal O}({\cal B}^4) \,,
\label{eq:magnetized-H}
\ee
where $J=k^\mu p_\mu$ is the conserved momentum associated with the generating Killing vector $k$. The leading correction thus consists of a term proportional to the conserved quantity $J^2$ and a position-dependent rescaling of the seed Hamiltonian.

Suppose the vacuum seed admits an independent quadratic Killing integral $K={\cal K}^{\mu\nu}p_\mu p_\nu$ satisfying the Poisson-bracket relation
\be
\{K,H\}=0 \,, \label{KHrelation}
\ee
where $\mathcal K^{\mu\nu}$ is invariant under the generating Killing vector, $\mathcal L_k\mathcal K^{\mu\nu}=0$, so that $\{K,J\}=0$. Testing the unchanged integral $\bar K=K$ in the transformed geometry gives
\be
\{\bar K,\bar H\}=-\frac{{\cal B}^2}{2}\{K,e^{-2\varphi}\}\bar H+{\cal O}({\cal B}^4) \,.
\label{eq:magnetized-ckt}
\ee
The bracket vanishes on the null shell through quadratic order. Hence, the seed KT itself defines a perturbative CKT of the magnetized metric, without requiring any correction to the quadratic integral $K$. For a stationary, axisymmetric seed with a complete Carter structure, provided $K$ is in involution with both Killing momenta, the corresponding null integrability and Hamilton-Jacobi separability are preserved through this order. On a timelike mass shell, the displayed term need not vanish, so this argument by itself does not establish a protected KT.

The existence of a CKT alone does not guarantee massless Klein-Gordon separability, since the associated second-order symmetry operator must also satisfy lower-order compatibility conditions. We address these conditions perturbatively by comparing the full scalar wave operators.

Since $k=\partial_z$ is a Killing vector, we can separate the $z$ dependence of the scalar field as $\bar\Phi=e^{i\ell z}\bar\phi$, with $\partial_z\bar\phi=0$. Using Eqs.~\eqref{eq:inverseg}--\eqref{eq:protection}, the scalar operator $\bar\Box=\frac{1}{\sqrt{-\bar g}}\partial_\mu(\sqrt{-\bar g}\,\bar g^{\mu\nu}\partial_\nu)$ satisfies
\be
\bar\Box\bar\Phi=
\left(1-\frac12{\cal B}^2e^{-2\varphi}\right)
(\Box -{\cal B}^2\ell^2)\bar\Phi
+{\cal O}({\cal B}^4) \,,
\label{eq:magnetized-KG}
\ee
where $\mu_{\rm eff}^2={\cal B}^2\ell^2$ acts as an effective squared mass in the seed equation. Consequently, a seed mode $\Phi$ with the same Fourier momentum $\ell$ satisfying
\be
(\Box-\mu_{\rm eff}^2)\Phi=0 \,, 
\ee
also solves the transformed massless equation through quadratic order upon setting $\bar\Phi=\Phi$. Complete massive Klein-Gordon separability of the seed, in a structure compatible with $k$, thus guarantees neutral massless Klein-Gordon separability after magnetization to this order. The effective mass is only an auxiliary parameter in the seed equation; the physical scalar on the magnetized background remains massless.


We now turn to inverse Harrison transformations, for which the resulting vacuum geometry may retain a deformation parameter $B$ inherited from the electrovacuum seed. We therefore allow a relation $B=B({\cal B})$, with $B(0)=0$, and regard both the vacuum geometry $g$ and its magnetized counterpart $\bar g$ as families parametrized by ${\cal B}$, sharing a common zero-${\cal B}$ limit. Equation~\eqref{eq:magnetized-H} then applies to each member of these families, with all quantities evaluated at the same ${\cal B}$.

Suppose the electrovacuum seed admits an independent quadratic conformal integral $\bar K=\bar{\cal K}^{\mu\nu}p_\mu p_\nu$ satisfying
\be
\{\bar K,\bar H\}=\bar\Lambda\bar H \,,
\ee
where CKT $\bar{\cal K}^{\mu\nu}$ ensures $\{\bar K,J\}=0$. We further assume that $\bar K$ admits a regular expansion $\bar K=K_0+{\cal O}({\cal B})$, with $K_0$ being a genuine Killing integral of the common zero-${\cal B}$ geometry. Since $\bar\Lambda$ vanishes in this limit, it can be expanded as
\be
\bar\Lambda={\cal B}\Lambda_1+{\cal B}^2\Lambda_2+{\cal O}({\cal B}^3) \,.
\ee
Identifying $K=\bar K$ as a candidate quadratic integral for the vacuum metric and using Eq.~\eqref{eq:magnetized-H}, we obtain
\be
\{K,H\}=
\left[
{\cal B}\Lambda_1+
{\cal B}^2\left(
\Lambda_2+\frac12\{K_0,e^{-2\varphi_0}\}
\right)
\right]H
+{\cal O}({\cal B}^3) \,,
\label{eq:inverse-CKT}
\ee
where $\varphi_0$ denotes the common zero-${\cal B}$ limit of $\varphi$. The demagnetized geometry therefore admits an independent CKT through order ${\cal B}^2$. If the inherited integral forms a regular, complete Carter-type separation structure compatible with both commuting Killing fields, the corresponding null Hamilton-Jacobi separability is also preserved through this order.


For the scalar equation, suppose the electrovacuum seed admits separation of the massless Klein-Gordon equation with a separation multiplier $q(r,x)$ and a field prefactor $w(r,x)$. After separating the Killing-coordinate dependence, the scalar operator takes the form
\be
q w^{-1}\bar\Box w
=\bar{\cal R}+\bar{\cal X} \,,
\label{eq:seed-KG-separation}
\ee
where $\bar{\cal R}$ and $\bar{\cal X}$ are the radial and angular operators, respectively. Regularity allows the expansion $q=q_0+{\cal O}({\cal B})$. If the common zero-${\cal B}$ geometry also admits massive Klein-Gordon separability with the same separation multiplier and field prefactor, then $q_0$ must be additively separable,
\be
q_0=q_r(r)+q_x(x) \,, 
\ee
which implies ${\cal B}^2\ell^2q={\cal B}^2\ell^2(q_r+q_x)+{\cal O}({\cal B}^3)$.
The inverse Harrison relation gives
\be
\left(1-\frac12{\cal B}^2e^{-2\varphi}\right)
\Box
=\bar\Box-{\cal B}^2\partial_z^2
+{\cal O}({\cal B}^3) \,.
\ee
Accordingly, for $\Phi=we^{-i\omega t+i\ell z}R(r)X(x)$, we obtain
\bea
&\!\!&\frac{q}{\Phi}
\left(1-\frac12{\cal B}^2e^{-2\varphi}\right)
\Box \Phi \nn\\
&=&\left[\frac{\bar{\cal R}R}{R}
+{\cal B}^2\ell^2q_r\right]
+\left[\frac{\bar{\cal X}X}{X}
+{\cal B}^2\ell^2q_x\right]
+{\cal O}({\cal B}^3) \,.
\label{eq:demag-KG-separation}
\eea
Since the two brackets depend only on $r$ and $x$, respectively, the massless scalar equation remains separable through quadratic order, yielding
\bea
\left(\bar{\cal R}
+{\cal B}^2\ell^2q_r\right)R
&=&\lambda R \,,\nn\\
\left(\bar{\cal X}
+{\cal B}^2\ell^2q_x\right)X
&=&-\lambda X \,,
\label{eq:demag-KG-radial-angular}
\eea
where $\lambda$ is a separation constant. The separated equations differ from those of the electrovacuum seed only at quadratic order. For regular local solutions with matching lower-order data, one can therefore construct $\Phi=\bar\Phi+{\cal B}^2\delta\Phi_2+{\cal O}({\cal B}^3)$ satisfying $\Box\Phi={\cal O}({\cal B}^3)$. Thus, massless Klein-Gordon separability is preserved through quadratic order under demagnetization.

Together, these results establish the quadratic-order inheritance of CKTs and massless scalar separability under magnetization and demagnetization, as summarized in Table~\ref{tab:inheritance}.

\begin{table}[t]
\caption{Perturbative inheritance of CKTs and massless Klein-Gordon (KG) separability under magnetic Harrison transformations.}
\label{tab:inheritance}
\begin{ruledtabular}
\begin{tabular}{lcc}
Transformation & Seed & Inherited \\
\hline
Magnetization
& KT & CKT \\
& Massive KG & Massless KG \\[3pt]
Demagnetization
& CKT & CKT \\
& Massless KG & Massless KG
\end{tabular}
\end{ruledtabular}
\end{table}


\textit{Obstructions to exact CKTs and massless scalar separability.}
The general perturbative inheritance established above leaves open a central question: does the inherited separability reflect an underlying exact separable structure, or can it persist even in geometries that are not exactly Carter-separable? Establishing the latter possibility requires a coordinate-independent obstruction to exact Carter-type separability, since failure of separation in a particular coordinate system does not rule out separability in another. We resolve this question in two representative geometries generated by magnetizing Kerr and demagnetizing KBR, and show that the latter possibility is realized in both cases.

As discussed above, complete Carter-type null separability requires an independent quadratic integral conserved on the null shell. The condition $\{K,H\}\propto H$ is equivalently expressed by the CKT equation
\be
\nabla_{(\mu}{\cal K}_{\nu\lambda)}=g_{(\mu\nu}{\cal V}_{\lambda)} \,,
\label{eq:ckt}
\ee
where ${\cal V}_\mu$ is an auxiliary one-form, and parentheses denote symmetrization over the enclosed indices. For the stationary and spacelike Killing vectors $\partial_t$ and $\partial_z$, the Carter integral must also be in involution with their conserved momenta $E=-p_t$ and $J=p_z$. The conditions $\{K,E\}=\{K,J\}=0$ imply that all ten components of ${\cal K}_{\mu\nu}$ depend only on $(r,x)$ in adapted coordinates, without assuming a block-diagonal form or preferred eigenvectors. The freedom ${\cal K}_{\mu\nu}\to{\cal K}_{\mu\nu}+fg_{\mu\nu}$, with arbitrary $f(r,x)$, leaves the null integral unchanged and preserves Eq.~\eqref{eq:ckt} under ${\cal V}_\mu\to{\cal V}_\mu+\nabla_\mu f$. We fix this freedom by imposing
\begin{equation}
g^{\mu\nu}{\cal K}_{\mu\nu}=0 \,.
\label{eq:tracefree}
\end{equation}
We retain all ten tensor components and impose Eq.~\eqref{eq:tracefree} as an additional linear constraint, giving fourteen unknown component functions: ten  of ${\cal K}_{\mu\nu}$ and four of ${\cal V}_\mu$. The Killing vectors $\partial_t$ and $\partial_z$ generate three reducible CKTs, corresponding to the quadratic integrals $E^2$, $EJ$, and $J^2$ on the null shell. An independent Carter constant therefore requires a fourth linearly independent CKT.

We employ the prolongation method~\cite{Frolov:2017kze,Branson:2007,Houri:2017tlk} to determine whether additional CKTs exist. At a regular point $(r_0,x_0)$, we differentiate Eq.~\eqref{eq:ckt} through total order $q$ and the trace-free condition~\eqref{eq:tracefree} through order $q+1$, with respect to $(r,x)$. Collecting the components of ${\cal K}_{\mu\nu}$ and their derivatives through order $q+1$, together with those of ${\cal V}_\mu$ through order $q$, into a vector $U_q$ gives a homogeneous linear system
\be
A_q U_q=0 \,,
\label{eq:matrixsystem}
\ee
where $A_q$ is an $R_q\times C_q$ matrix, with $R_q=(q+2)(21q+23)/2$ and $C_q=(q+2)(7q+17)$. Every symmetry-invariant trace-free CKT must satisfy this system, while the three reducible CKTs ensure $\operatorname{nullity}(A_q)\geq3$. A nullity of three would exclude any additional independent CKT at the corresponding order.

To determine the rank, we evaluate $A_q$ at a representative regular point with rational parameter values, $(m,a,{\cal B},r_0,x_0)=(2,1,20/101,5,1/2)$, and perform exact Gaussian elimination over the finite field $\mathbb F_p$, with $p=1000003$. All denominators remain nonzero modulo $p$. This procedure provides an exact lower bound on the rank for generic parameter values. Combined with the three known reducible CKTs, a computed nullity of three therefore excludes any additional independent CKT. As shown in Table~\ref{tab:ranks}, both magnetized Kerr and demagnetized KBR reach nullity three at $q=2$, which remains unchanged at $q=3$. Since every CKT must satisfy the prolonged equations, these results show that only the three reducible CKTs survive locally for generic parameter values. In contrast, the Kerr limit ${\cal B}=0$ has nullity four at both orders, recovering the Carter symmetry. The absence of a fourth independent CKT therefore excludes an additional quadratic Carter constant and exact Carter-type null separability.

\begin{table}[t]
\caption{
Nullities of the CKT system at different prolongation orders $q$, computed modulo $p=1000003$ at the chosen checkpoint. 
}
\label{tab:ranks}
\begin{ruledtabular}
\begin{tabular}{lcccc}
$q$               & 0  & 1  & 2   & 3   \\
\hline
$R_q$             & 23 & 66 & 130 & 215 \\
$C_q$             & 34 & 72 & 124 & 190 \\
Magnetized Kerr   & 11 & 6  & 3   & 3   \\
Demagnetized KBR  & 11 & 6  & 3   & 3   \\
Kerr              & 11 & 6  & 4   & 4
\end{tabular}
\end{ruledtabular}
\end{table}

The obstruction also extends to complete Carter-type $R$-separation of the massless Klein-Gordon equation~\cite{Gray:2025lwy,Michel:2013dfa}. Such separation requires an independent second-order symmetry operator whose principal symbol defines a CKT, even in the presence of a nontrivial scalar prefactor. The absence of an additional CKT therefore excludes exact Carter-type massless scalar separability in any coordinate system compatible with the two Killing symmetries.


\textit{Leading-order obstruction to CKY inheritance.}
The preceding results establish that CKTs and massless scalar separability are inherited at leading nontrivial order in ${\cal B}$ from (conformal) Carter-type seed geometries under (de)magnetization, despite their exact breakdown. This naturally raises the question: does the (C)KY structure associated with first-order Dirac symmetry operators exhibit the same perturbative inheritance? Although Petrov type I rules out exact CKY two-forms in these geometries~\cite{Glass:1998ka,Batista:2014fpa}, it does not by itself determine the order at which CKY inheritance fails. We show that the seed-connected (C)KY structure is already obstructed at leading nontrivial order in both representative cases.

To establish this perturbative obstruction, we examine the eigenvalue structure of the self-dual Weyl tensor. In four dimensions, the Weyl tensor can be represented by a complex $3\times3$ matrix $W$ acting on self-dual two-forms. For a Petrov type-D seed, this matrix takes the diagonal form $W_0=\mathrm{diag}(-2\lambda_0,\lambda_0,\lambda_0)$ in a suitable basis, where $\lambda_0$ is the repeated eigenvalue. Splitting this degeneracy violates a necessary algebraic condition for the existence of a CKY two-form, thereby obstructing the inheritance of the seed CKY structure.

To determine whether such splitting occurs, we expand $W=W_0+{\cal B}W_1+{\cal B}^2W_2+\cdots$. For both magnetized Kerr and demagnetized KBR, the metrics are even in ${\cal B}$, so $W_1=0$ in a suitable regular frame. In the eigenbasis of $W_0$, let $V$ denote the lower-right $2\times2$ block of $W_2$, corresponding to the two equal seed eigenvalues. Expanding the eigenvalue equation gives $\lambda_\pm=\lambda_0+{\cal B}^2\mu_\pm+{\cal O}({\cal B}^4)$, where $\lambda_\pm$ are the perturbed Weyl eigenvalues and $\mu_\pm$ are the eigenvalues of $V$. The squared difference between these eigenvalue corrections is
\begin{equation}
\delta_2\equiv(\mu_+-\mu_-)^2
=2\tr(V^2)-(\tr V)^2 \,.
\label{eq:split}
\end{equation}
Thus, $\delta_2\neq0$ establishes Weyl eigenvalue splitting at quadratic order, obstructing the perturbative inheritance of the seed CKY structure.

Starting from the quadratic metric expansion in Eq.~\eqref{eq:inverseg}, the leading Weyl eigenvalue splitting under magnetization can be determined directly from a Ricci-flat type-D seed and its generating Killing vector, without constructing the full magnetized metric. For demagnetization, the type-D structure of the electrovacuum family requires the Weyl eigenvalue splitting of the resulting vacuum geometry to be canceled by the forward Harrison transformation at the same order.

For magnetized Kerr and demagnetized KBR, the common zero-field limit is Kerr. Evaluating the quadratic Weyl corrections for the corresponding generating Killing vectors, we obtain, in both cases,
\begin{equation}
\delta_2
=\frac{9\Delta^2(1-x^2)^2}{4(r+iax)^4} \,,
\label{eq:cky-splitting}
\end{equation}
where $x=\cos\theta$ and $\Delta=r^2-2mr+a^2$. This expression is nonzero at generic regular points, demonstrating that the seed-connected CKY structure is already obstructed at ${\cal O}({\cal B}^2)$ in both geometries, while CKTs and massless scalar separability survive through the same order.


\textit{Discussion.}
The preceding analysis reveals a hierarchy in the survival of hidden symmetries under magnetic Harrison transformations, combining general perturbative inheritance laws with explicit obstructions established for magnetized Kerr and demagnetized KBR, as summarized in Table~\ref{tab:hierarchy}. Small departures from Kerr are often of particular physical interest, including those induced by a weak external magnetic field. In this regime, the inherited separability provides an analytically tractable framework for studying leading corrections to photon trajectories and massless scalar-wave propagation. The loss of exact Carter-type separability also motivates a search for chaotic photon dynamics beyond this separable approximation~\cite{Wang:2021ara}. Moreover, the inheritance laws follow directly from the magnetic Harrison transformation and the seed separation structure, without requiring the explicit form of each generated metric. They therefore provide a practical means of identifying separable approximations in more complicated generated geometries.

\begin{table}[t]
\caption{
Hierarchy of hidden symmetries and scalar separability under magnetic Harrison transformations.
}
\label{tab:hierarchy}
\begin{ruledtabular}
\begin{tabular}{lccc}
$\quad$ & CKT & Massless scalar separability & CKY \\
\hline
Exact
& $\times$ & $\times$ & $\times$ \\
${\cal O}({\cal B}^2)$ 
& $\checkmark$ & $\checkmark$ & $\times$
\end{tabular}
\end{ruledtabular}
\end{table}

Although the exact CKT and leading-order CKY obstructions have been established here for magnetized Kerr and demagnetized KBR, the coordinate-independent methods are not restricted to these geometries. We conjecture that the loss of exact Carter-type separability is a generic feature of rotating geometries generated by magnetic Harrison transformations and their inverses. This conjecture can be tested in broader families, including demagnetized genuine KBR~\cite{Ovcharenko:2026tos,Ma:2026gdn} and magnetized Kerr-Taub-NUT~\cite{Siahaan:2021uqo}. Establishing general obstruction criteria directly from the Harrison transformation would provide a deeper understanding of the origin and scope of this apparent universality.

More broadly, recent developments in constructing new Petrov type D seed geometries describing black holes immersed in external magnetic fields~\cite{Podolsky:2025tle,Ovcharenko:2026tos,Ovcharenko:2025cpm,Furugori:2026qbb}, together with extensions of magnetic solution-generation techniques to nonlinear electrodynamics~\cite{Bokulic:2025ucy} and Kaluza-Klein and Einstein-Maxwell-dilaton theories~\cite{Ma:2026rro}, provide broader opportunities for applying and extending the hidden-symmetry inheritance laws established here. These developments also motivate investigating analogous universal inheritance laws in supergravity and exploring whether similar general perturbative inheritance laws exist under other solution-generating transformations, such as string dualities~\cite{Chervonyi:2015ima}.


\textit{Acknowledgments.}
We are grateful to  H. L\"u for useful discussions. 
S.L. is supported in part by the National Natural Science Foundation of China (No. 12105098, No. 12481540179) and the Natural Science Foundation of Hunan Province (No. 2022JJ40264), and by the Excellent Young Scholars Program of the Hunan Provincial Department of Education under Grant No. 25B0092. 
H-D.L.~is  supported in part by National Natural Science Foundation of China Grants No.12447134 and Postdoctoral Innovation Project of Shandong Province SDCX-ZG-202503036.  




%

\end{document}